# Optimized Implementation of Memristor-Based Full Adder by Material Implication Logic

Mehri Teimoory*, Amirali Amirsoleimani‡, Jafar Shamsi**, Arash Ahmadi†, Shahpour Alirezaee‡, Majid Ahmadi‡

Email: {mehri_teimoory}@yahoo.com {amirsol,alirezae,ahmadi}@uwindsor.ca {jafarshamsi}@elec.iust.ac.ir {ahmadi}@razi.ac.ir

*Department of Electrical Engineering, Islamic Azad University, Science and Research Branch, Kermanshah, Iran

‡Department of Electrical and Computer Engineering, University of Windsor, Windsor, Ontario, Canada

**Department of Electrical Engineering, Iran University of Science and Technology, Tehran, Iran

†Department of Electrical Engineering, Razi University, Kermanshah, Iran

***Abstract*— Recently memristor-based applications and circuits are receiving an increased attention. Furthermore, memristors are also applied in logic circuit design. Material implication logic is one of the main areas with memristors. In this paper an optimized memristor-based full adder design by material implication logic is presented. This design needs 27 memristors and less area in comparison with typical CMOS-based 8-bit full adders. Also the presented full adder needs only 184 computational steps which enhance former full adder design speed by 20 percent.**

## I. Introduction

Logic computing has recently drawn wide attention due to emergence of memory-resistor or memristor devices [1]. Memristor's non-volatility good scalability, low power consumption and compatibility with CMOS structures makes them an ideal device for various applications.

Logic operations with memristors are promising for nano computing circuits. Several approaches are presented for implementing logic circuits with memristors. Hybrid CMOS-memristor based logic and stand-alone memristor based logic are two main logic family [1]. Hybrid CMOS-Memristor based logic is based on integrating CMOS with memristors only for computational purpose. Memristor Ratioed Logic (MRL) is presented in [2]. In this family, logical states represented in voltage logic gates based on MRL are not capable to store their last output value. The other drawback in MRL design is its need to the conventional CMOS inverter gate for building the NAND and NOR gates. The stand-alone memristor based logics are more promising than hybrid peer. Crossbar architecture is also one of the family logic with memristor presented in [3], where the computation takes place in crossbar structure of memristors. Since in this family of logic the fabrication is simple and its high density make it to be a good candidate for future computing architectures. Other type of innovative memristor-based logic gates are developed and presented in [4] and [5], where the resistance represents logical state unlike MRL. Therefore they can store output of the gates as memristance of the output memristor. Both of these families can perform functions inside a crossbar array but they need a sequencer to bias memristors. Material implication which is known IMPLY is presented in [4][5]. This logic family executes basic Boolean operations by applying memristor based IMPLY logic gate.

In this paper a serial eight-bit memristor based full adder is implemented based on material implication (IMPLY) logic. This full adder is designed with 27 memristors and it needs 184 computational steps for completing its operation which is faster than previous designs [6-7].

This paper is organized as follows. Section II introduces IMPLY logic gate and its utilization in building different logic gates. Design and serial implementation of one-bit memristor based full adder as well as the proposed 8-bit adder are presented in section III. Consequently the section IV includes the conclusion of the paper.

## II. Material Implication (IMPLY) Logic

$p$ IMPLY $q$ is a logic function that is called material implication. This logic statement means "$p$ implies $q$" or "if $p$ then $q$". In Table 1 a truth table of material implication logic is presented. This logic is similar to answering a question. $p$ is considered as a question and $q$ as its answer. When $p$ is wrong any answer results in a true output ( logic 1) whether it is wrong or correct one. In other cases when p is true the answer truthfulness would determine output. Only case three of IMPLY logic truth table as it is shown in Table 1 has FALSE (logic 0) output. In this case a correct question ($p$ = 1) has wrong answer ($q$ = 0). Based on the output values for each cases of IMPLY logic this function is the equivalent to,

$$p \quad \text{IMP} \quad q \cong p \rightarrow q \cong (\sim p) \vee q \qquad (1)$$

TABLE. 1. Truth Table Of IMPLY Function.

| Case | $p$ | $q$ | $p$ IMP $q$ |
|---|---|---|---|
| **1** | 0 | 0 | 1 |
| **2** | 0 | 1 | 1 |
| **3** | 1 | 0 | 0 |
| **4** | 1 | 1 | 1 |

### A. IMPLY logic gate

The memristor-based IMPLY logic gate constructed by two memristor ($P$ and $Q$) and one resistor ($R_G$). The polarity of the memristors is depicted in Fig. 1. Also the $R_G$ connects the memristors into the ground. Two voltage biases for IMPLY logic gate, $V_{cond}$ and $V_{set}$ are imposed into $P$ and $Q$ memristors doped side respectively for $P$ IMP $Q$ operation. The $V_{cond}$ voltage magnitude is lower than $V_{set}$ ($|V_{cond}| < |V_{set}|$). In this design the proposed IMPLY gate inputs are stored as resistance of the $P$ and $Q$ memristors. Interestingly, output of this gate is the final changed memristance value of $Q$ memristor. For setting input of IMPLY logic gate initial memristances of this memristors should be set based on the considered input values of each case showed in Table. 1. For this purpose two bias sources are required to set initial memristance of memristors to $R_{ON}$ and $R_{OFF}$. For implementing logic zero memristors memristance should be in its highest value $R_{OFF}$. The memristor should be inversely biased. Therefore $V_{clear}$ voltage source is applied to set memristors initial memristances to OFF state. On the other hand, for setting logic one with memristor, its memristance should be changed to its lowest value $R_{ON}$. $V_{set}$ voltage source can be used for forward bias of the memristor. It should be mentioned both of these bias voltages should be higher than threshold voltage of the memristors. Also pulse width of the proposed voltage should be long enough for reaching the memrisance of the memristor to the considered values ($R_{ON}$ or $R_{OFF}$). For different input combinations stated in Table 1 IMPLY logic gate produces different outputs. In case 1, the memristances of $P$ and $Q$ should be both set to $R_{OFF}$ initially. $R_G$, $R_{ON}$ and $R_{OFF}$ are assumed 10 kΩ, 1 kΩ and 100 kΩ respectively. For this both memristor are biased with $V_{clear}$, *which* is considered -1 V. Then $V_{cond}$ and $V_{set}$ biases are connected to $P$ and $Q$ memristors respectively. For simulation $V_{cond}$ and $V_{set}$ are 0.5 V and 1 V respectively. By imposing these voltages based on voltage division between $R_G$ ($R_G$ = 10 kΩ) and resistance of memristor $Q$ ($R_{OFF}$ = 100 kΩ), voltage of the node between two memristor and $R_G$ becomes negligible. The voltage across memristor $Q$ is,

$$V_Q = \frac{R_G}{R_{OFF} + 2R_G}\left(V_{set} + V_{cond}\right) \tag{2}$$

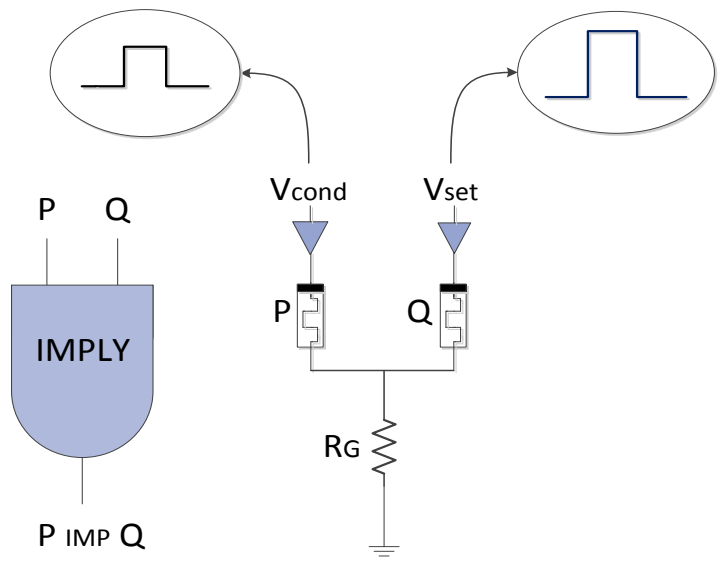

Figure 1: Memristor-based IMPLY logic gate.

This makes current flows from $V_{set}$ to ground. As the charge passes through memristor $Q$ its memristance is decreasing until it reaches to its minimum value $R_{ON}$. Therefore the final memristance ($R_{ON}$) of $Q$ is the desired output (logic 1) of the IMPLY gate for case 1 of the truth table. The case 1 is simulated for IMPLY gate with mentioned characteristic in Fig. 2. This case is determined the write time of the circuit.

In case 2, the memristances of $P$ and $Q$ should be set to $R_{OFF}$ and $R_{ON}$ respectively. This would be done by biasing $P$ and $Q$ memristors with $V_{clear}$ and $V_{set}$ respectively to allocate desired input to memristors.

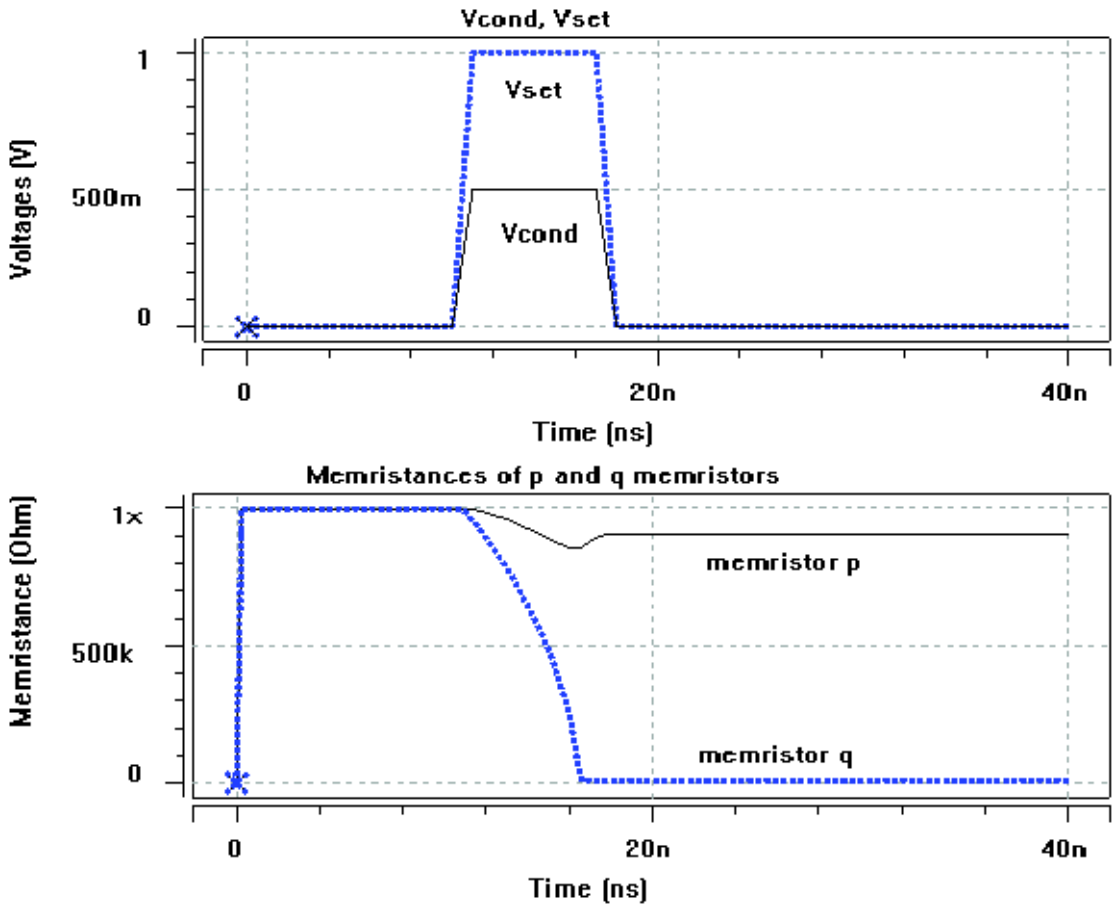

Figure 2: The simulations of IMPLY gate for case 1. Linear ion drift model is applied for memristors.

Then $V_{cond}$ and $V_{set}$ voltage bias sources should connect to $P$ and $Q$ to start IMPLY logic operation. Since memristance of memristor $Q$ ($R_{ON}$ = 1 kΩ) is negligible in comparison with $R_G$ resistance ($R_G$ = 10 kΩ), in this case voltage of the common node of memristors and resistor would be about 1 V. The voltage across memristor $Q$ in this case is determined through,

$$V_Q = V_{set} - \frac{R_{ON}}{R_{OFF}}\frac{R_{OFF} + R_G}{R_{ON} + 2R_G}V_{set} \cong 0 \tag{3}$$

Therefore the voltage across memristor $Q$ would be zero. As a result, the memristor $Q$ state is remained unchanged as its previous state (logic 1). Case 4 is similar to this case. Its only difference with case 2 is setting procedure for inputs of memristors. The voltage across memristor $Q$ is determined by,

$$V_Q = \frac{R_G}{R_{ON} + 2R_G}\left(V_{set} + V_{cond}\right) \tag{4}$$

In case 3, the memristance of $Q$ is set to $R_{OFF}$. After applying $V_{cond}$ and $V_{set}$ voltage bias sources the logic state remain unchanged. The voltage across the memristor $Q$ is,

$$V_Q = \frac{R_G}{R_{ON} + R_G}V_{cond} \cong V_{cond} \tag{5}$$

This voltage is not enough for changing the state of memristor. Although this voltage is forced the internal state of $Q$ to change to $R_{ON}$ (logic 1) it cannot be changed to its minimum value. This is called state drift. In this state final memristance of $Q$ becomes lower than initial input memristance ($R_{OFF}$).

### B. *Implementing logic functions by IMPLY logic*

By combining IMPLY logic gate and FALSE gate any logic structure can be produced [4-5]. FALSE gate is a logic gate which yields zero as its output for any input. This circuit is consisting of three memristors for implementing NAND logic based on IMPLY and FALSE functions. $P$ and $Q$ are inputs and $S$ is output. The NAND function for $P$ and $Q$ is written by IMPLY and FALSE logic through,

$$S = Q \text{ IMP } (P \text{ IMP } S) = \sim Q \vee ((\sim P) \vee S) \tag{6}$$

By allocating zero to $S$ then

$$S = \sim Q \vee ((\sim P) \vee 0) = (\sim Q) \vee (\sim P) = \sim (P \wedge Q) \quad (7)$$

This shows for the first step FALSE operation should be applied to memristor *S,* which means $V_{clear}$ should be applied to memristor *S* (*S* IMP 0). For implementing *P* IMP *S,* voltage pulses of $V_{cond}$ and $V_{set}$ are applied to memeristors *P* and *S* respectively. Final step is producing output *S* = (*Q* IMP *S*). For this step $V_{cond}$ and $V_{set}$ are applied to memeristors *Q* and *S* respectively. The memristance of *Q* is considered as output of this logic. The truth table for each step is shown in Table 2 for NAND logic gate. Similarly other logic gates can be implemented based on this procedure. Different logic operations made through IMPLY logic and FALSE function is introduced in Table 3.

## III. 8-Bit Memristor Based Full Adder Based On Material Implication (IMPLY) Logic

In CMOS based 8-bit full adder two eight-bit inputs are entered and one eight-bit output number is extracted. Single CMOS 8-bit full adder comprises of 400 CMOS transistors. Several implementations were applied for designing 8-bit full adder based on material implication logic. The proposed method in [6] has used 712 computational steps and it also needs 29 memristors for implementing an 8-bit full adder. In [7] 27 memristors are used for implementing an 8-bit full adder with 232 computational steps. The proposed method is based on architecture which is depicted in Fig. 3, where the output of the proposed full adder is determined by,

$$S = A \oplus B \oplus C_{in} \quad (8)$$

$$C_{out} = (A \cdot B) + (C_{in} \cdot (A \oplus B)) \quad (9)$$

where *A*, *B* are inputs and $C_{in}$ is carry in for one bit full adder. Also $C_{out}$ and *S* are carry out and output of the full adder. In this design XOR, AND and OR logics should be implemented by material implication logic. For implementing XOR function unlike typical implementation of XOR logic, as mentioned in Table 3, two optimized equivalent IMPLY based XOR designs are proposed. These implementations are developed based on design requirements of the proposed serial full adders. These implementations have decreased computational steps of XOR logic. The proposed IMPLY based XOR implementations are,

$$A \oplus B = (A \text{ IMP } B) \text{ IMP } (A' \text{ IMP } B')' \quad (10)$$

*A* XOR *B*: FALSE($M_0$), *A* IMP $M_0$, FALSE($M_1$), *B* IMP $M_1$, *A* IMP *B*, $M_0$ IMP $M_1$, FALSE($M_0$), $M_1$ IMP $M_0$, *B* IMP $M_0$

It takes 9 computational steps.

$$A \oplus B = ((A' \text{ IMP } B) \text{ IMP } (A \text{ IMP } B'))' \quad (11)$$

*A* XOR *B*: FALSE($M_0$), *A* IMP $M_0$, FALSE($M_1$), *B* IMP $M_1$, $M_0$ IMP *B*, *A* IMP $M_1$, FALSE($M_0$), $M_1$ IMP $M_0$, *B* IMP $M_0$, FALSE($M_1$), $M_0$ IMP $M_1$

It takes 11 computational steps. As it can be seen Eqn. 10 and Eqn. 11 need nine and eleven computational steps. In typical implementation in the first steps input memristors (*A* and *B*) are used as inputs of IMPLY logic. This will change their memristances and logic values. Therefore these values should be copied in other memristors for using in subsequent steps. This procedure will result in additional computational steps. In Eqn. 10 and Eqn. 11 by changing IMPLY based XOR logic design there is no need for keeping input values since $A'$ and $B'$ are required in subsequent steps. Therefore computational steps are reduced. Using this implementation one bit full adder is designed.

TABLE. 2. NAND gate Implementation Truth Table Based IMPLY Function.

| Step 1: FALSE (*S*) | Step 2: *P* IMP *S* | | | Step 3: *Q* IMP *S* | | |
|---|---|---|---|---|---|---|
| ***S*** | ***P*** | ***S*** | ***S'*** | ***Q*** | ***S*** | ***S''*** |
| 0 | 0 | 0 | 1 | 0 | 1 | 1 |
| 0 | 0 | 0 | 1 | 1 | 1 | 1 |
| 0 | 1 | 0 | 0 | 0 | 0 | 1 |
| 0 | 1 | 0 | 0 | 1 | 0 | 0 |

TABLE. 3. Logic Gates Implementation Based on IMPLY Function.

| Logic Gate | Implementation based on IMPLY |
|---|---|
| **NOT *P*** | *P* IMP 0 |
| ***P* NAND *Q*** | *P* IMP (*Q* IMP 0) |
| ***P* AND *Q*** | {*P* IMP (*Q* IMP 0)} IMP 0 |
| ***P* NOR *Q*** | {(*P* IMP 0) IMP *Q*} IMP 0 |
| ***P* OR *Q*** | (*P* IMP 0) IMP *Q* |
| ***P* XOR *Q*** | (*P* IMP *Q*) IMP {(*Q* IMP *P*) IMP 0} |

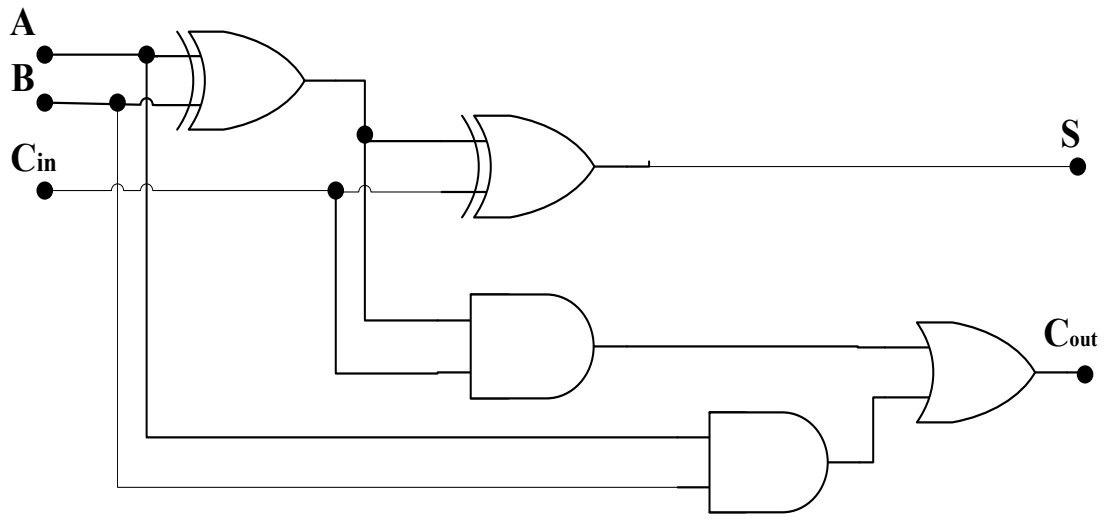

Figure 3: Logic schematic of the proposed design for eight-bit full adder.

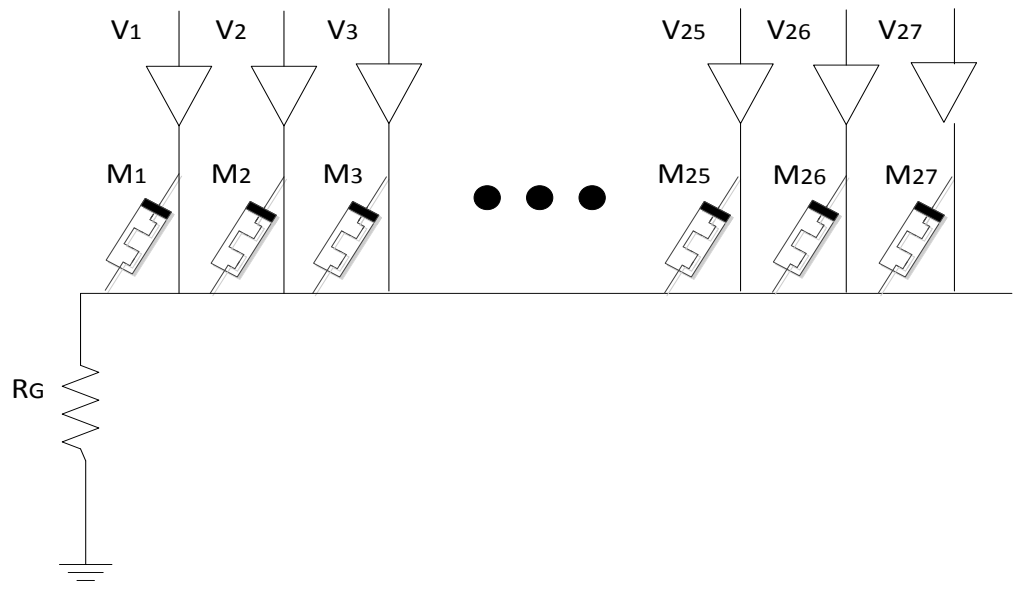

Figure 4: Proposed memristor-based eight-bit full adder.

This full adder needs only 23 computational steps by applying 6 memristors. *A* and *B* are used as input memristors. $M_1$ and $M_2$ are applied as functional devices. Also *C* is used as both input and output. Carry in ($C_{in}$) value is stored in memristor *C* and its last value becomes carry out ($C_{out}$). *S* is the last value stored in $M_3$. For designing N-bit full adder with the proposed design, input memristors of previous determined bits are used as functional memristors for computing purpose of the next bits. This will decrease number of memristors dramatically. Also carry in of each bit is the last value of memristor *C*. Therefore for $C_{in}$ and $C_{out}$ only one memristor is used for all bits of N-bit full adder. By this technique except first bit which needs six memristors other bits only required three memristors. For implementing 8-bit full adder 27 memristors are used in a row as it depicted in fig. 4. This full adder completes its task in 184 computational steps. It lowers 48 steps in comparison with 8-bit full adder designed in [7]. In Table 4 and Table 5 logics and input and output values are defined in each step.

## IV. CONCLUSION

Material implication logic is a significant issue in digital system design. In this paper, an optimized serial implementation technique for 8-bit memristor based full adder is presented. This technique takes 23 computational steps for one-bit full adder. By using 27 memristors an 8-bit full adder is designed, which requires 184 computational steps to complete its task. This shows about 20 percent faster performance in comparison with the techniques presented in [6] and [7].

TABLE. 4. VALUES AND LOGICS IN EACH STEP FOR PROPOSED DESIGN ONE-BIT FULLADDER.

| Step | Goal | FALSE and IMPLY Logic in each step | Parameters Value and Equivalent Logic | Output |
|---|---|---|---|---|
| 0 | Setting Initial Values | - | *A* and *B* and *C* is set to initial values | |
| 1 | | FALSE ($M_1$) | $M_1 = 0$ | |
| 2 | | FALSE ($M_2$) | $M_2 = 0$ | |
| 3 | | FALSE ($M_3$) | $M_3 = 0$ | |
| 4 | Copy *A*' to $M_2$ | $A \rightarrow M_3$ | $A' = M_3$ | |
| 5 | Copy *B*' to $M_2$ | $B \rightarrow M_2$ | $B' = M_2$ | |
| 6 | (*A* XOR *B*) | $M_3 \rightarrow B$ | *A'* IMP *B* | |
| 7 | | $A \rightarrow M_2$ | *A* IMP *B*' | |
| 8 | | $M_2 \rightarrow M_1$ | (*A* IMP *B*') IMP 0 | |
| 9 | | $B \rightarrow M_1$ | $M_1$=(*A* XOR *B*)' | |
| 10 | | FALSE (*A*) | $A = 0$ | |
| 11 | | $M_1 \rightarrow A$ | *A* = (*A* XOR *B*) | |
| 12 | | FALSE (*B*) | $B = 0$ | |
| 13 | | FALSE ($M_3$) | $M_3 = 0$ | |
| 14 | | $C \rightarrow M_3$ | *C*' | |
| 15 | | $M_3 \rightarrow A$ | *C*' IMP (*A* XOR *B*) | |
| 16 | | $C \rightarrow M_1$ | *C* IMP (*A* XOR *B*)' | |
| 17 | | $M_1 \rightarrow B$ | (*C* IMP (*A* XOR *B*)')' | |
| 18 | | $A \rightarrow B$ | *S* ' | |
| 19 | | FALSE ($M_3$) | $M_3 = 0$ | |
| 20 | Finish *S* execution | $B \rightarrow M_3$ | $M_3$= (*A* XOR *B* XOR *C*) | ***S*** |
| 21 | | FALSE (*C*) | $C = 0$ | |
| 22 | | $M_2 \rightarrow C$ | | |
| 23 | Finish $C_{out}$ execution | $M_1 \rightarrow C$ | $C = C_{out}$ | $C_{out}$ |

TABLE. 5. TRUTH TABLE OF MEMRISTOR-BASED 1-BIT FULL ADDER BASED ON IMPLY FUNCTION.

| MEMRISTORS | *A* | | | | *B* | | | | | *C* | | | | $M_1$ | | | | $M_2$ | | | $M_3$ | | | | | |
|---|---|---|---|---|---|---|---|---|---|---|---|---|---|---|---|---|---|---|---|---|---|---|---|---|---|---|
| INPUTS / OUTPUTS | INPUT 1 | | | | INPUT 2 | | | | | $C_{IN}$ | | | $C_{OUT}$ | | | | | | | | | | | | | *S* |
| STEPS | 0 | 10 | 11 | 15 | 0 | 6 | 12 | 17 | 18 | 0 | 21 | 22 | <u>23</u> | 1 | 8 | 9 | 16 | 2 | 5 | 7 | 3 | 4 | 13 | 14 | 19 | <u>20</u> |
| CASE 1 | 0 | 0 | 0 | 0 | 0 | 0 | 0 | 0 | 1 | 0 | 0 | 0 | 0 | 0 | 0 | 1 | 1 | 0 | 1 | 1 | 0 | 1 | 0 | 1 | 0 | 0 |
| CASE 2 | 0 | 0 | 0 | 1 | 0 | 0 | 0 | 0 | 0 | 1 | 0 | 0 | 0 | 0 | 0 | 1 | 1 | 0 | 1 | 1 | 0 | 1 | 0 | 0 | 0 | 1 |
| CASE 3 | 0 | 0 | 1 | 1 | 1 | 1 | 0 | 0 | 0 | 0 | 0 | 0 | 0 | 0 | 0 | 0 | 1 | 0 | 0 | 1 | 0 | 1 | 0 | 1 | 0 | 1 |
| CASE 4 | 0 | 0 | 1 | 1 | 1 | 1 | 0 | 1 | 1 | 1 | 0 | 0 | 1 | 0 | 0 | 0 | 0 | 0 | 0 | 1 | 0 | 1 | 0 | 0 | 0 | 0 |
| CASE 5 | 1 | 0 | 1 | 1 | 0 | 1 | 0 | 0 | 0 | 0 | 0 | 0 | 0 | 0 | 0 | 0 | 1 | 0 | 1 | 1 | 0 | 0 | 0 | 1 | 0 | 1 |
| CASE 6 | 1 | 0 | 1 | 1 | 0 | 1 | 0 | 1 | 1 | 1 | 0 | 0 | 1 | 0 | 0 | 0 | 0 | 0 | 1 | 1 | 0 | 0 | 0 | 0 | 0 | 0 |
| CASE 7 | 1 | 0 | 0 | 0 | 1 | 1 | 0 | 0 | 1 | 0 | 0 | 1 | 1 | 0 | 1 | 1 | 1 | 0 | 0 | 0 | 0 | 0 | 0 | 1 | 0 | 0 |
| CASE 8 | 1 | 0 | 0 | 1 | 1 | 1 | 0 | 0 | 0 | 1 | 0 | 1 | 1 | 0 | 1 | 1 | 1 | 0 | 0 | 0 | 0 | 0 | 0 | 0 | 0 | 1 |